\documentclass{article}

\usepackage{arxiv}
\usepackage{amsmath,amsfonts}
\usepackage{algorithm}
\usepackage{algpseudocode}
\usepackage{array}
\usepackage{textcomp}
\usepackage{stfloats}
\usepackage{url}
\usepackage{verbatim}
\usepackage{graphicx}
\usepackage{subcaption}
\usepackage{cite}
\usepackage{listings}
\usepackage{xspace}
\usepackage[hidelinks]{hyperref}
\usepackage{multirow}
\usepackage{color}

\usepackage{draftwatermark}
\SetWatermarkText{PREPRINT}
\SetWatermarkLightness{0.95}

\newif\iftechr
\techrtrue

\newcommand{\gthree}{Graviton3~(c7g)\xspace}
\newcommand{\gfour}{Graviton4~(c8g)\xspace}

\newcommand{\xeon}{Xeon~(c7i)\xspace}
\newcommand{\yitian}{Yitian~(c8y)\xspace}
\newcommand{\amd}{EPYC~(c7a)\xspace}
\newcommand{\axion}{Axion~(c4a)\xspace}

\newcommand{\prover}{zkProver\xspace}

\date{}

\begin{document}

\title{On the Performance of Cloud-based ARM SVE for Zero-Knowledge Proving Systems}

\iftechr
\author{Dumitrel Loghin, Shuang Liang, Shengwei Liu, Xiong Liu, Pingcheng Ruan, Zhigang Ye \\
\\
        OKX Group}
\fi



\maketitle

\begin{abstract}

Zero-knowledge proofs (ZKP) are becoming a gold standard in scaling blockchains and bringing Web3 to life. At the same time, ZKP for transactions running on the Ethereum Virtual Machine require powerful servers with hundreds of CPU cores. The current zkProver implementation from Polygon is optimized for x86-64 CPUs by vectorizing key operations, such as Merkle tree building with Poseidon hashes over the Goldilocks field, with Advanced Vector Extensions (AVX and AVX512). With these optimizations, a ZKP for a batch of transactions is generated in less than two minutes. With the advent of cloud servers with ARM which are at least 10\% cheaper than x86-64 servers and the implementation of ARM Scalable Vector Extension (SVE), we wonder if ARM servers can take over their x86-64 counterparts. Unfortunately, our analysis shows that current ARM CPUs are not a match for their x86-64 competitors. Graviton4 from Amazon Web Services (AWS) and Axion from Google Cloud Platform (GCP) are 1.6X and 1.4X slower compared to the latest AMD EPYC and Intel Xeon servers from AWS with AVX and AVX512, respectively, when building a Merkle tree with over four million leaves. This low performance is due to (1) smaller vector size in these ARM CPUs (128 bits versus 512 bits in AVX512) and (2) lower clock frequency. On the other hand, ARM SVE/SVE2 Instruction Set Architecture (ISA) is at least as powerful as AVX/AVX512 but more flexible. Moreover, we estimate that increasing the vector size to 512 bits will enable higher performance in ARM CPUs compared to their x86-64 counterparts while maintaining their price advantage.

\end{abstract}

\section{Introduction}

ARM-based servers have entered the cloud computing market and offer attractive pricing compared to traditional x86-64 servers equipped with Intel and AMD CPUs. For example, ARM-based servers offered by Amazon Web Services (AWS) are around 20\% cheaper compared to their x86-64 counterparts and they can deliver up to 40\% higher performance~\cite{aws_arm}. Currently, cloud providers such as AWS, Alibaba Cloud, Microsoft Azure, Google Cloud Platform (GCP), and Huawei Cloud, among others, offer cloud servers (also known as instances or Virtual Machines -- VMs) equipped with ARM CPUs~\cite{dumi_tcc2024}. Among these servers, we show in a previous study~\cite{dumi_tcc2024} that \textit{c7g} family from AWS equipped with Graviton3 and \textit{c8y} family from Alibaba Cloud equipped with Yitian 710 offer the best performance among the existing cloud ARM servers.

Scalable Vector Extension (SVE and SVE2)~\cite{svepaper} are novel features of ARM CPUs, being part of the 64-bit ARMv8-A and ARMv9-A Instruction Set Architectures (ISA). These extensions bring ARM in direct competition with x86-64 CPUs equipped with Advanced Vector Extensions (AVX, AVX2, and AVX512). In our previous paper~\cite{dumi_tcc2024}, we show that existing data management software platforms do not make use of SVE. In contrast, in this paper we analyze the performance of ARM-based servers with SVE and SVE2 when running Zero-Knowledge Proof (ZKP) systems, and in particular the Poseidon hash~\cite{poseidon_hash} over the 64-bit Goldilocks field~\cite{goldilocks} which are key components of the Polygon \prover~\cite{zkprover_doc}.

ZKP technologies are currently used by Layer 2 (L2) blockchains in the Web3 space providing fast settlement and privacy for L2 transactions through ZK rollups~\cite{zk_okx}. For example, Polygon's zkEVM~\cite{zkevm_doc} is an Ethereum L2 ZK rollup solution that offers good compatibility with the Ethereum Virtual Machine (EVM) while producing ZKPs of the correct execution of transactions on the L2 EVM. These ZKPs are then uploaded to Ethereum (also referred to as Layer 1 or L1) and verified by a smart contract. X Layer~\cite{xlayer} is an L2 based on Polygon's zkEVM where the \prover~\cite{zkprover_doc}, the component that produces the ZKP for a batch of transactions, runs on cloud-based instances with x86-64 CPUs with large numbers of cores and a significant amount of RAM. For example, a \textit{m7a.48xlarge} server from AWS with a 192-core AMD EPYC CPU at 3.7~GHz and 768~GB of RAM produces a proof for 300 transactions in 90 seconds. Such an excessive amount of cloud resources translate into significant financial burden. For example, the above-mentioned instance costs around \$11 per hour when used on demand or \$60,000 per year when reserved for one year. 

ARM-based cloud instances are typically cheaper than x86-64-based instances~\cite{aws_arm, dumi_tcc2024}. However, \prover is a highly optimized software that uses AVX2 and AVX512 vectorization on x86-64 servers to produce a ZKP in less than two minutes. To test the suitability of ARM servers, we first need to overcome the challenge of implementing \prover with ARM SVE support and then benchmark its performance on cloud-based servers. Since there is a limited number of research studies on ARM SVE, in the process of implementing \prover with ARM SVE we need to assess the suitability and effectiveness of using this ISA to implement cryptographic primitives. For benchmarking, we have limited options: only AWS Graviton3, AWS Graviton4, Alibaba Yitian 710, and Goolge Axion support SVE. Graviton3 supports 256-bit registers but implements only SVE, while Graviton4, Yitian 710, and Axion implement SVE2 but have only 128-bit registers. At the moment, there is no ARM CPU with 512-bit SVE registers, hence, it is impossible to fairly compare the performance of an ARM SVE/SVE2 implementation against its x86-64 AVX512 counterpart.

In summary, we make the following contributions and key observations in this paper:
\begin{itemize}
    \item We implement Goldilocks field operations, Poseidon hashing, and Merkle tree building with ARM SVE and SVE2 support. Our implementation supports vectors of 128, 256, and 512 bits in length and is open source\footnote{Code available at \url{https://github.com/okx/goldilocks/tree/dev-sve}.} (more details in Section~\ref{sec:impl}).
    \item We argue that SVE and SVE2 ISAs are not inferior to AVX2 and AVX512 ISA families. We find implementing Goldilocks field operations easier and more efficient with ARM SVE/SVE2 assembly and intrinsics (more details in Section~\ref{sec:impl}).
    \item We perform a comprehensive performance analysis on four types of ARM CPUs available in the cloud, namely, Graviton3 and Graviton4 from AWS, Axion from GCP, and Yitian 710 from Alibaba Cloud. We show that Axion and Graviton4 achieve the highest performance among the ARM CPUs when building a Merkle tree with $2^{22}$ leaves using Poseidon hashes over the Goldilocks field. However, they are still $1.4\times$ and $1.6\times$ slower compared to Intel and AMD CPUs with AVX512. This lower performance boils down to the following facts: (1) smaller vector size of 128 bits in Axion and Graviton4 compared to 512 bits in x86-64 CPUs with AVX512, and (2) lower clock frequency (2.8 and 3.1 GHz in Graviton4 and Axion, respectively, compared to 3.7 GHz in Intel Xeon and AMD EPYC CPUs).
\end{itemize}

We hope that our findings will guide ARM hardware designers and cloud providers in improving SVE support in the next few years. Firstly, ARM hardware designers should consider implementing SVE execution units with larger registers (at least 512 bits) and increasing the clock frequency to compete with Intel and AMD CPUs. In Section~\ref{sec:dis}, we estimate that increasing the vector size to 512 bits will enable higher performance in ARM CPUs compared to their x86-64 counterparts, while the impact on power and cost of this performance increase is not as significant as reducing the competitive advantage of ARM-based cloud instances. Secondly, cloud providers should consider fully utilizing the hardware capabilities of existing ARM CPUs, as we show that Graviton3 and Graviton4 are underclocked. In the remainder of this paper, we present the details of our analysis.

\section{Background}

In this section, we introduce the context of our study for readers who are not familiar with vectorization, blockchains, L2, and ZKP systems.

\subsection{Vectorization, AVX, and SVE}

Vectorization (via vector operations) is a technique for improving the performance of data-parallel applications on modern CPUs by applying the same instruction to multiple data instances simultaneously. Hence, vectorization is a type of Single Instruction, Multiple Data (SIMD) parallelism based on Flynn's taxonomy~\cite{flynn}. This technique makes use of hardware support in modern CPUs for vector operations. For example, x86-64 CPUs from Intel and AMD support Advanced Vector Extension (AVX), AVX2, and AVX512 (where AVX512 has multiple separate instruction sets such as AVX512 Foundation (F) and AVX512 Doubleword and Quadword Instructions (DQ), among others)~\cite{avx_book}. 

AVX and AVX2 work with 256-bit registers being able to pack 8-, 16-, 32-, and 64-bit integers, as well as, 32- and 64-bit IEEE 754 floating points into one register. For example, four 64-bit integers can be processed at the same time by one AVX2\footnote{AVX2 adds support for integers on top of the floating point supported by AVX. In this paper, we use the term \textbf{AVX} to refer to AVX plus AVX2.} instruction. As suggested by its name, AVX512 operates on 512-bit registers, being able to pack twice the number of operands compared to AVX. In this paper, we use both AVX2 and AVX512 Foundation (F) instruction sets.

SVE is the alternative offered by ARM for vectorization. It is supposed to replace ARM's previous vectorization extension called NEON.  While NEON operates on fixed 128-bit registers, SVE supports any vector length in the range of 128 to 2048 bits, in steps of 128 bits~\cite{introsve}. This ISA does not encode the vector length, thus, enabling the developers to write generic code that can run on different register sizes. However, the basic packed types are the same: 8-, 16-, 32-, and 64-bit signed and unsigned integers, as well as, 32- and 64-bit IEEE 754 floating points. It is worth noting that SVE supports unsigned integer operations, while AVX only operates with signed integers. SVE2~\cite{introsve2} includes all SVE instructions while adding a few more instructions to enhance the implementation of workloads outside HPC and AI domains. We shall see in Section~\ref{sec:impl} that SVE2 allows us to implement Goldilocks multiplication with only two instructions compared to 18 in SVE.

%

\subsection{Ethereum and L2}

Ethereum~\cite{Ethereum_2013} was the first blockchain~\cite{rpc_book} to support smart contracts by providing a Turing-complete execution environment called the Ethereum Virtual Machine (EVM). Currently, Ethereum is the second blockchain in the world in terms of total market value, after Bitcoin~\cite{Bitcoin_2008}, with a total market value of \$460 billion as of May 2024. By supporting smart contracts, Ethereum is one of the main choices of Web3 developers who implement decentralized apps (dApps) using smart contracts.

A smart contract is a computer program consisting of functions and variables, resembling the concept of objects and classes in object-oriented programming. A smart contract is executed on multiple nodes (or peers) and it is supposed to produce the same results (deterministic execution) on all these nodes. Public functions in smart contracts can be called by other contracts or by users via Externally Owned Accounts (EOA). A public function that modifies a variable (or state) that is stored on the blockchain is processed via a transaction that is included in the blockchain ledger and it is executed by all the blockchain nodes.

In Ethereum, a smart contract function call (transaction) has an associated fee called gas fee which is expressed in GigaWei (GWei)\footnote{1 Wei = $10^{-18}$ ETH. 1 GWei = $10^{-9}$ ETH.}, a denomination of Ether (ETH), the native cryptocurrency of Ethereum. This fee has multiple roles. First, the gas fees of the transactions included in a block are distributed to the miner (or block proposer) that added the block to the chain (we direct the reader to~\cite{rpc_book} for more details). Second, the fee deters malicious users from sending spam transactions to attack the blockchain (e.g., performing denial of service). However, when many honest users compete in including their transactions in the blockchain and the price of ETH is high, the transaction fees skyrocket\footnote{A chart of historic gas fee price can be accessed at \url{https://etherscan.io/chart/gasprice}.}, making it difficult for developers and companies to implement new smart contracts.

In addition to the above issue, Ethereum also suffers from low scalability and low throughput~\cite{rpc_book}. To address these issues, the Ethereum community introduced Layer 2 (L2) networks where users can deploy their smart contracts while the Ethereum blockchain (also called Layer 1 or L1) ensures the security of an L2. An L2 has its own fee mechanism; in practice, L2 fees are much smaller than those on Ethereum. Moreover, there can virtually be an infinity of L2 networks that connect to a single L1, thus improving the scalability of Ethereum.

L2s come in two flavors, namely optimistic rollups and ZK rollups. In optimistic rollups, L2 transactions are submitted to L1, and users can check their correctness during a challenge period. After this challenge period, the transactions are considered final. In contrast, ZK rollups produce a cryptographic proof that L2 transactions are valid and this proof is automatically checked in a smart contract on L1. ZK rollups can also protect the privacy of the transaction data and have much faster finality since there is no need for a challenge period. However, ZK rollups require powerful hardware to compute the complex ZK proofs. In the next sections, we look at ZK proofs and ZK provers in detail.

\subsection{X Layer and \prover}

\begin{figure}[tbp]
\centering
\includegraphics[width=0.7\textwidth]{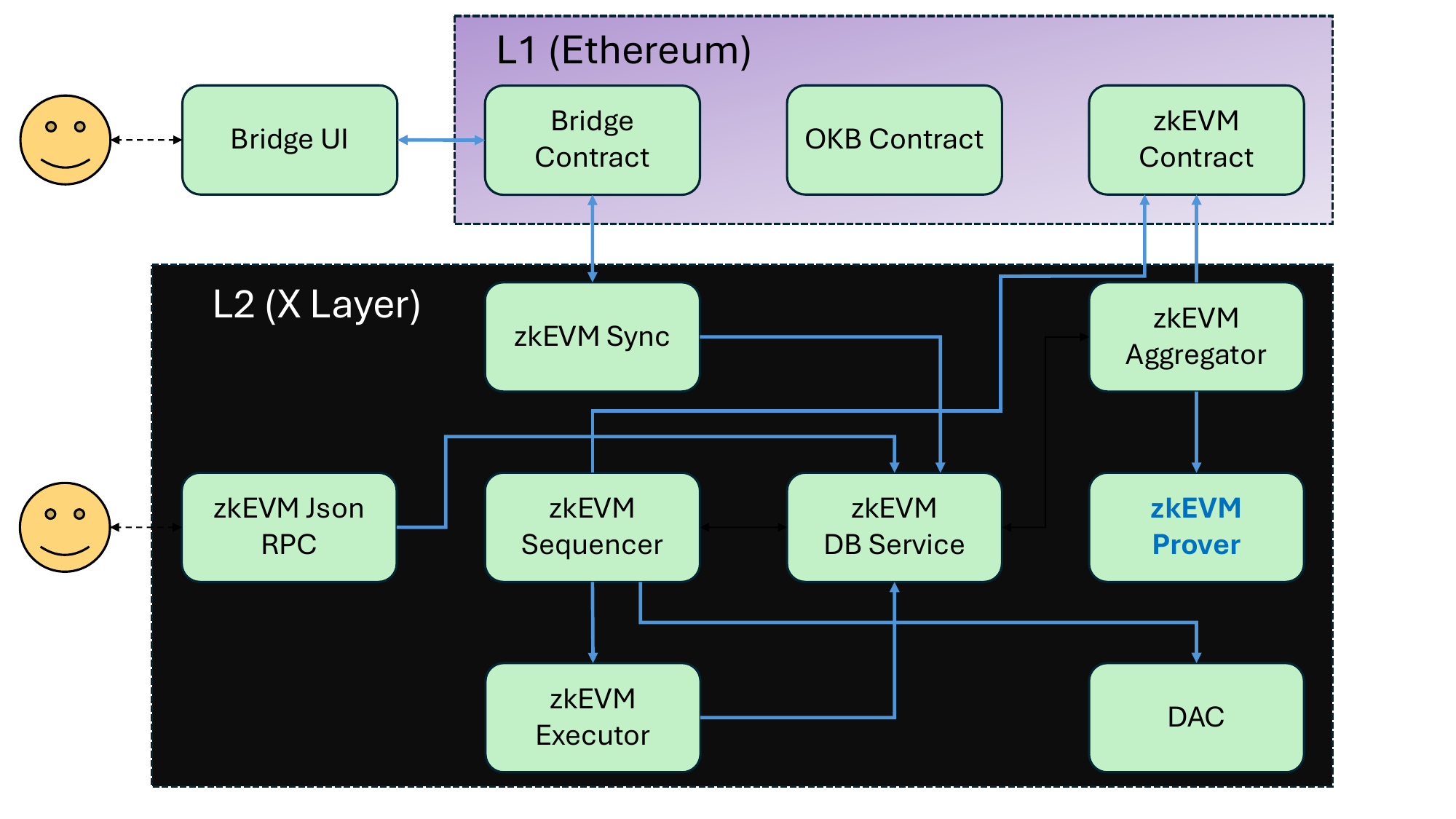}
\caption{X Layer overview with \prover highlighted.}
\label{fig:xlayer_arch}
\end{figure}

X Layer~\cite{xlayer} is an L2 for Ethereum operated by the OKX Group and built on the Polygon Chain Development Kit (CDK)~\cite{cdk_doc}. As such, X Layer is a ZK rollup L2 with an architecture described in Fig.~\ref{fig:xlayer_arch}. A user interacts with the L2 via the JSON RPC service where the user can send queries and transactions. The RPC service saves these requests to a database operated by the DB Service. The Sequencer takes the transactions from the database and packs them in blocks and the blocks into batches. The Executor executes these transactions following the EVM specifications and updates the state database (which is also operated by the DB Service). The Aggregator takes the transactions batches and asks the Prover (henceforth \prover) to produce a ZKP per batch. The Aggregator then uploads the proof to L1 via the zkEVM Smart Contract. This contract verifies the validity of the proof. When a user wants to transfer her funds from L1 to L2, she interacts with the Bridge Smart Contract on L1. An L2 user needs to pay the gas fees on X Layer in OKB, operated by the OKB Smart Contract, which is of ERC-20 type.

\prover has a key role in the X Layer and the Polygon CDK. \prover is described in~\cite{zkprover_paper} and implemented in C++ with AVX2 and AVX512 optimizations\footnote{X Layer version: \url{https://github.com/okx/xlayer-prover}. \\ Polygon version: \url{https://github.com/0xPolygonHermez/zkevm-prover}.}. A key component of \prover is the Merkle tree builder for polynomial commitments. This builder uses Poseidon hashing~\cite{poseidon_hash} over Goldilocks field elements~\cite{goldilocks}. The code for this component is in the \texttt{goldilocks} repository\footnote{X Layer version: \url{https://github.com/okx/goldilocks}. \\ Polygon version: \url{https://github.com/0xPolygonHermez/goldilocks}.}. Our performance profiling of \prover shows that it spends close to 30\% of its execution time in Merkle tree building, 10\% in Number Theoretic Transform (NTT), and 5\% in polynomial evaluations, all over the Goldilocks field. Hence, improving the performance of Goldilocks field operations leads to significant improvements in all three hotspots.


\subsection{Zero Knowledge Proofs}

Informally, a ZKP's role is to convince a \textit{verifier} about the truth of a \textit{prover}'s claim, without the prover revealing unnecessary information. For example, a prover wants to convince a verifier that she knows the password of her social media account without revealing the password to the verifier. In the context of ZK rollups, a ZKP needs to convince the smart contract verifier on L1 that the execution of a batch of transactions on L2 was done following the EVM specifications and that these transactions are valid (e.g., there is no double-spending).

\begin{algorithm}[tp]
\caption{Pseudo-code of Poseidon hash permutation.}
\label{alg:perm}
\begin{algorithmic}
\State $\textbf{Function}~\texttt{SBox(Goldilocks x)}$
\Return $x^7$
\\
\State $\textbf{Function}~\texttt{Add(Goldilocks x, Goldilocks y)}$ \\
\Return $x + y$
\\
\State $\textbf{Function}~\texttt{MDS(Goldilocks state[12],}$
\State $\texttt{Goldilocks mat[12][12])}$
\Return $state \times mat$
\\
\State $\textbf{Function}~\texttt{MDSPartial(Goldilocks state[12],}$
\State $\texttt{Goldilocks vec[2][12])}$
\State $tmp \gets state \cdot vec[0]$
\State $W \gets state[0] \times vec[1]$ \Comment{$W$ is a vector ($W[12]$)}
\State $state \gets Add(state, W)$
\State $state[0] \gets tmp$ \\
\Return $state$
\\
\State $\textbf{Function}~\texttt{Permute(Goldilocks state[12])}$
\State $state \gets Add(state, C)$\Comment{$C, M, S$ are vectors of constants}
\For{$HALF\_FULL\_ROUNDS$}
    \State $state \gets SBox(state)$
    \State $state \gets MDS(state, M)$
    \State $state \gets Add(state, C)$
\EndFor
\For{$PARTIAL\_ROUNDS$}
    \State $state[0] \gets SBox(state[0])$
    \State $state \gets MDSPartial(state, S)$
\EndFor
\For{$HALF\_FULL\_ROUNDS$}
    \State $state \gets SBox(state)$
    \State $state \gets MDS(state, M)$
    \State $state \gets Add(state, C)$
\EndFor \\
\Return $state$
\end{algorithmic}
\end{algorithm}

\textbf{Goldilocks Field.} \prover represents the EVM execution with a series of trace polynomials with coefficients in the Goldilocks field. This field is a finite field determined by the Goldilocks prime 
\begin{equation}
\label{eq:gp}
P = 2^{64} - 2^{32} + 1
\end{equation}
This prime has the advantage that it can be represented on 64 bits and, thus, the operations in the field can be efficiently executed on modern CPUs (that have 64-bit registers). For more details, we refer the reader to~\cite{goldilocks} and ~\cite{gp_blog}. In this paper, we refer to a Goldilocks element $e$ as canonical if and only if $0 \leq e < P$.

\textbf{Merkle Tree.} \prover constructs a series of Merkle trees over polynomials' evaluations to commit to these polynomials. A Merkle tree is a binary tree where the leaf layer consists of hashes (e.g., Poseidon hashes) of the target data elements (e.g., transactions in Ethereum or polynomial evaluations in \prover). In the next layers up to the root, each node is the hash of its children hashes concatenation. Thus, the root of a Merkle tree uniquely represents the underlying leaves (that is, the change of a single bit in the leaves renders a completely different root hash). \prover deals with Merkle trees of varying sizes, where the largest has more than 80~GB at leaf level. Due to these large size trees, a server needs more than 350~GB RAM to run the \prover.

\textbf{Poseidon Hash.} To build a Merkle tree, \prover uses the Poseidon hash~\cite{poseidon_hash} over the Goldilocks field. Poseidon is a sponge-based hash with an internal state consisting of 12 Goldilocks elements. This sponge "absorbs" a variable-length input in chunks (by default, chunks of eight Goldilocks elements) by applying a permutation function over the internal state. This permutation function is the computationally heavy part of Poseidon. Algorithm~\ref{alg:perm} presents Poseidon's permutation function. It proceeds in several rounds, where $HALF\_FULL\_ROUNDS = 4$ and $PARTIAL\_ROUNDS = 22$. A full round firstly applies $SBox()$ consisting of raising each state element to the seventh power in the Goldilocks field. Hence, each $SBox()$ consists of multiplications and reductions such that the result is a valid Goldilocks field element. Secondly, a full round applies an $MDS()$ operation which is a vector-matrix multiplication between the state vector and a matrix of constants. This matrix is also called the maximum distance separable (MDS) matrix. Lastly, some round constants are added to the state. A partial round consists of applying the $SBox()$ only to the first state element, followed by dispersing the state elements among each other via scalar and vector products combined with some constants (note that $M, C,$ and $S$ in Algorithm~\ref{alg:perm} are vectors and matrices of constants). In summary, Poseidon hash in \prover consists of Goldilocks field operations such as addition, multiplication, and reduction. In the next section, we present the details of these operations and their implementation using AVX and SVE vector operations.

\section{Implementation}
\label{sec:impl}

In this section, we first describe the existing AVX and AVX512 implementations of the Poseidon hash, followed by the particularities of our SVE and SVE2 implementation.

\subsection{Existing AVX Implementation} 

Since the Poseidon sponge consists of 12 Goldilocks elements (64 bits each), three AVX registers (256 bits each) are needed to represent the sponge. In the AVX512 implementation, three AVX512 registers represent two sponges, such that two hashes are computed simultaneously. The \texttt{SBox()} and \texttt{Add()} operations are straight-forward, however, \texttt{SBox()} implies multiplications that need reductions. When two 64-bit elements are multiplied, they produce a 128-bit result. This result needs to be reduced back to a Goldilocks element.

\begin{algorithm}[tp]
\caption{Pseudo-code of Goldilocks reduction of a 128-bit unsigned integer.}
\label{alg:reduce}
\begin{algorithmic}
\State $\textbf{Function}~\texttt{Reduce128(128-bit integer r)}$
\State $r_l = r \And 0xFFFFFFFF$ \Comment{$\And$ is bitwise AND}
\State $r_h = r >> 64$ \Comment{$>>$ is shift right}
\State $r_{hl} = r_h \And 0xFFFF$
\State $r_{hh} = r_h >> 32$
\State $tmp1 = r_l - r_{hh}$
\If{$r_{hh} > r_l$}
\State $tmp1 = tmp1 - 2^{32} + 1$   \Comment{fix underflow}
\EndIf
\State $tmp2 = (2^{32}-1) * r_{hl}$ \Comment{64-bit}
\State $res = tmp1 + tmp2$
\If{$tmp1 > 2^{64} - 1 - tmp2$}
\State $res = res + 2^{32} - 1$ \Comment{this is always $<P$}
\ElsIf{$res \geq P$}
\State $res = res - P$
\EndIf \\
\Return $res$
\end{algorithmic}
\end{algorithm}

Given two Goldilocks elements $a$ and $b$, their product is $r = a \cdot b$, where $r < 2^{128}$. This result can be expressed as
\begin{equation}
    r = r_{h} \cdot 2^{64} + r_{l} = (r_{hh} \cdot 2^{32} + r_{hl}) \cdot 2^{64} + r_l
\end{equation}
where $r_h < 2^{64}$, $r_l < 2^{64}$, $r_{hh} < 2^{32}$, and $r_{hl} < 2^{32}$. In the Goldilocks field with the prime $P$ defined in Equation~\ref{eq:gp},
\begin{equation}
    2^{64} = P + 2^{32} - 1 \iff 2^{64} \equiv 2^{32} - 1 (mod P)
\end{equation}
Hence, 
\begin{equation}
\begin{aligned}
    r \equiv (r_{hh} \cdot 2^{32} + r_{hl}) \cdot (2^{32} - 1) + r_l \\ = r_{hl} \cdot (2^{32} - 1) + r_l - r_{hh} \\ \leq (2^{32}-1)\cdot(2^{32} -1) + 2^{64} - 1 = 2 \cdot P - 2 < 2 \cdot P
\end{aligned}
\end{equation}
This reduction can be implemented as depicted in Algorithm~\ref{alg:reduce}. The AVX implementation follows the same steps depicted in Algorithm~\ref{alg:reduce}, with the difference that each step is applied to four 64-bit elements. This reduction is implemented by a function called \texttt{reduce\_avx\_128\_64()} consisting of 12 AVX instructions.

\begin{algorithm}[tp]
\caption{Pseudo-code of Goldilocks Add() in AVX using signed integers.}
\label{alg:addavx}
\begin{algorithmic}
\State $\textbf{Function}~\texttt{Add(Goldilocks a, Goldilocks b)}$
\State $a_s \gets shift(a)~\textbf{where}~a_s = a~XOR~2^{63}$
\State $a_{sc} \gets to\_canonical(a_s)$ 
\State $c_s \gets a_{sc} + b$
\If{$a_{sc} > c_s$}
\State $c_s \gets c_s - P$
\EndIf
\State $c \gets shift(c_s)~\textbf{where}~c = c_s~XOR~2^{63}$ \\
\Return $c$
\end{algorithmic}
\end{algorithm}
\begin{algorithm}[bp]
\caption{Pseudo-code of Goldilocks Add() in SVE using unsigned integers.}
\label{alg:addsve}
\begin{algorithmic}
\State $\textbf{Function}~\texttt{Add(Goldilocks a, Goldilocks b)}$
\State $c \gets a + b$
\State $d \gets P - a$
\If{$b >= d$}
\State $c \gets c - P$
\EndIf \\
\Return $c$
\end{algorithmic}
\end{algorithm}

The \texttt{Add()} operation is straightforward but AVX does not support unsigned 64-bit integers, hence, all the sub-operations are done with signed 64-bit integers. With 64-bit signed integers, we have many cases to check since there is a hardware overflow at $2^{63}$ and a field overflow at $P$. To better check for the Goldilocks field overflow, the existing AVX implementation "shifts" one operand by $2^{63}$ (via a $XOR$ instruction), as shown in Algorithm~\ref{alg:addavx}. In addition, the code checks if the input is a valid Goldilocks element or converts it via the \texttt{to\_canonical()} function. Briefly, this function subtracts $P$ from the input if the input is greater than or equal to $P$. The assumption is that the input cannot exceed $2\cdot P$. In total, there are 15 AVX assembly instructions per Goldilocks \texttt{Add()} as shown in code Listing~\ref{lstaddavx} (if we exclude data loading/off-loading, there are 9 assembly instructions). In contrast, we implemented this operation with 10 (or 6 if we exclude data loading/off-loading instructions) SVE assembly instructions, as shown in code Listing~\ref{lstaddsve}.

\begin{figure}[tp]
\centering
\begin{minipage}{0.45\textwidth}
\begin{lstlisting}[caption={Goldilocks $Add(a, b)$ in AVX Assembly},label=lstaddavx,basicstyle=\ttfamily\small,numbers=left]
vmovdqa   %3, %%ymm13 // MSB
vmovdqa   %4, %%ymm14 // P_s
vmovdqa   %5, %%ymm15 // P_n
vmovdqa   %1, %%ymm11 // a
vmovdqa   %2, %%ymm12 // b
vpxor     %%ymm11, %%ymm13, %%ymm11
vpcmpgtq  %%ymm14, %%ymm11, %%ymm10
vpandn    %%ymm10, %%ymm15, %%ymm10
vpaddq    %%ymm10, %%ymm11, %%ymm10
vpaddq    %%ymm10, %%ymm12, %%ymm11
vpcmpgtq  %%ymm11, %%ymm10, %%ymm14
vpand     %%ymm14, %%ymm15, %%ymm10
vpaddq    %%ymm10, %%ymm11, %%ymm10
vpxor     %%ymm10, %%ymm13, %%ymm10
vmovdqa   %%ymm10, %0
\end{lstlisting}
\end{minipage}
\quad\quad\quad\quad
\begin{minipage}{0.45\textwidth}
\vspace{-45pt}
\begin{lstlisting}[caption={Goldilocks $Add(a, b)$ in SVE Assembly},label=lstaddsve,basicstyle=\ttfamily\small,numbers=left]
ptrue   p7.b, all
ld1d    z31.d, p7/z, %1     // a
ld1d    z29.d, p7/z, %2     // b
mov     z28.d, #4294967295  // P_n
mov     z30.d, #-4294967295 // P
sub     z30.d, z30.d, z31.d
add     z31.d, z29.d, z31.d
cmphi   p6.d, p7/z, z29.d, z30.d
add     z31.d, p6/m, z31.d, z28.d
st1d    z31.d, p7, %0
\end{lstlisting}
\end{minipage}
\begin{minipage}{0.5\textwidth}
\begin{lstlisting}[caption={$Add(a, b)$ in SVE Assembly as generated by GCC 13 ($a$ and $b$ are in $z0.d$ and $z3.d$, respectively)},label=lstaddsvegcc,basicstyle=\ttfamily\small,numbers=left]
ptrue   p0.b, all
mov     z1.d, #4294967295
movprfx z4.d, p0/z, z3.d
add     z4.d, p0/m, z4.d, z0.d
mov     z2.d, #-4294967295
movprfx z2.d, p0/z, z2.d
sub     z2.d, p0/m, z2.d, z0.d
movprfx z0.d, p0/z, z4.d
add     z0.d, p0/m, z0.d, z1.d
cmphi   p0.d, p0/z, z3.d, z2.d
sel     z0.d, p0, z0.d, z4.d
\end{lstlisting}
\end{minipage}
\end{figure}

Similar to \texttt{Add()}, \texttt{Sub()} starts by shifting $a$, but it also shifts $b$ and puts it into canonical form. Next, the actual subtraction is performed. If $b$ is greater than $a$, the subtraction underflows, hence, it is transformed back into canonical form by adding $P$. Since both operands are shifted, there is no need to perform a shifting of the result. In summary, \texttt{Sub()} needs the same number of AVX instructions as \texttt{Add()}.

\begin{algorithm}[tp]
\caption{Pseudo-code of Goldilocks Sub() in AVX using signed integers.}
\label{alg:subavx}
\begin{algorithmic}
\State $\textbf{Function}~\texttt{Sub(Goldilocks a, Goldilocks b)}$
\State $a_s \gets shift(a)~\textbf{where}~a_s = a~XOR~2^{63}$
\State $b_s \gets shift(b)~\textbf{where}~b_s = b~XOR~2^{63}$
\State $b_{sc} \gets to\_canonical(b_s)$
\State $c \gets a_{s} - b_{sc}$
\If{$b_{sc} > a_s$}
\State $c \gets c + P$
\EndIf \\
\Return $c$
\end{algorithmic}
\end{algorithm}

\begin{algorithm}[tp]
\caption{Pseudo-code of Goldilocks Sub() in SVE using unsigned integers.}
\label{alg:subsve}
\begin{algorithmic}
\State $\textbf{Function}~\texttt{Sub(Goldilocks a, Goldilocks b)}$
\State $c \gets a - b$
\If{$a < b$}
\State $c \gets c + P$
\EndIf \\
\Return $c$
\end{algorithmic}
\end{algorithm}

\begin{figure}[tp]
\centering
\includegraphics[width=0.7\textwidth]{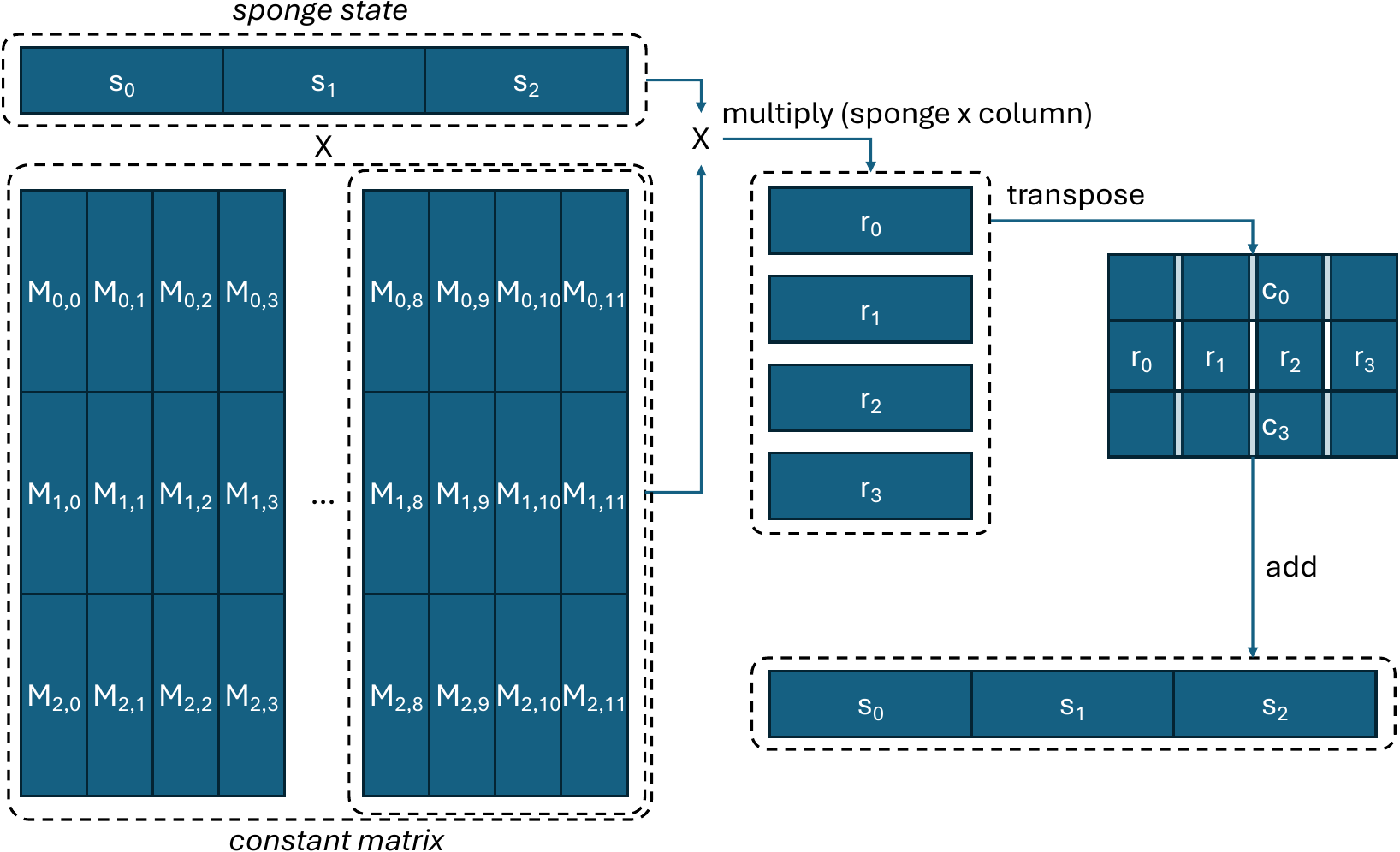}
\caption{A visualization of the MDS() function.}
\label{fig:mds}
\end{figure}

Lastly, we discuss the \texttt{MDS()} and \texttt{MDSPartial()} functions which consist of vector-matrix multiplications. \texttt{MDS()} multiplies the sponge which is a vector of 12 Goldilocks elements with an MDS matrix of size $12 \times 12$. The intuition behind this function is to disperse state changes throughout the entire sponge. In its vectorized implementation, the sponge is split into three AVX registers of four elements each, and the matrix into three sub-matrices of size $12 \times 4$ (where each of the 12 columns is stored in three AVX registers), as shown in Fig.~\ref{fig:mds}. The sponge is multiplied by each of these three sub-matrices. For each sub-matrix, the multiplication leads to four rows of four elements each, depicted by $r_{0}, r_{1}, r_{2},$ and $r_{3}$ in Fig.~\ref{fig:mds}. Next, all four elements in a row need to be added in the Goldilocks field. To perform these additions, we need to transpose the rows and use \texttt{Add()} to get the final result consisting of four Goldilocks elements which represent the new state of the sponge (e.g., $s_2$ in Fig.~\ref{fig:mds}). In summary, \texttt{MDS()} makes use of all the basic functions in the AVX implementation, as we shall see below.

\begin{figure}[tp]
\centering
\begin{subfigure}{0.8\textwidth}
\includegraphics[width=1.0\textwidth]{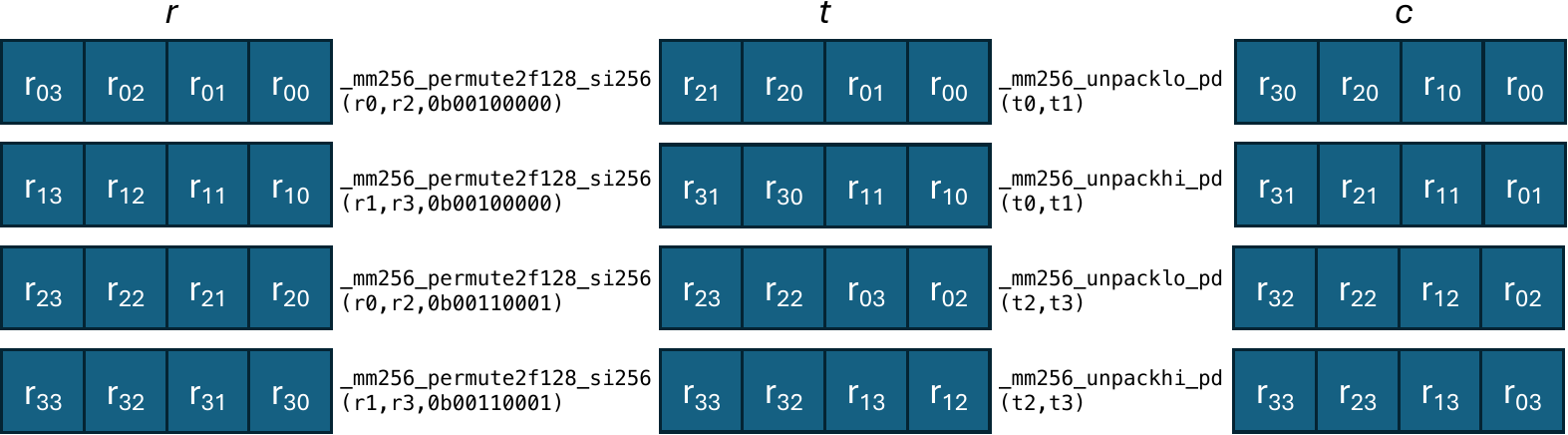}
\caption{In AVX.}
\label{fig:transavx}
\end{subfigure}
\quad
\begin{subfigure}{0.8\textwidth}
\includegraphics[width=1.0\textwidth]{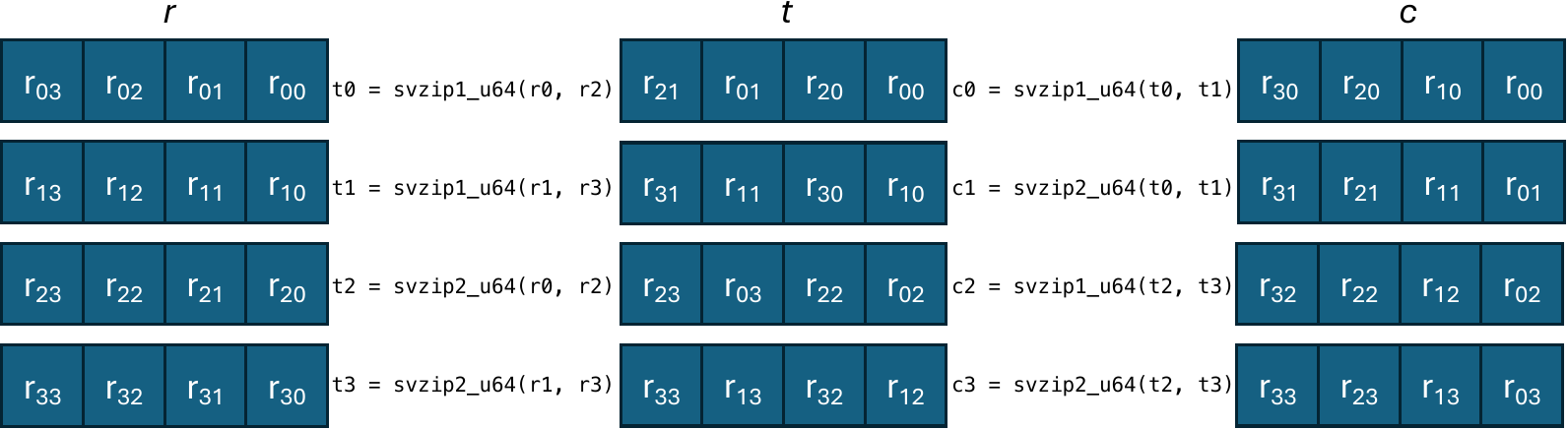}
\caption{In SVE.}
\label{fig:transsve}
\end{subfigure}
\caption{A visualization of the transpose code.}
\label{fig:trans}
\end{figure}

It is worth discussing the transposition shown in Fig.~\ref{fig:mds} which is part of \texttt{MDS()}. This is done using specialized instructions in both AVX and SVE. A visualization of the transpose codes in AVX and SVE is presented in Fig.~\ref{fig:trans}. In AVX, we first use \texttt{\_mm256\_permute2f128\_si256(a,b,c)} to get intermediary registers where the first 128 bits come from parameter $a$ and the next 128 bits come from parameter $b$. The exact portion of the bits to be copied is determined by the value of $c$. We present the values of $a, b,$ and $c$ in Fig.~\ref{fig:transavx}. Next, \texttt{\_mm256\_unpacklo\_pd(a,b)} and \texttt{\_mm256\_unpackhi\_pd(a,b)} are used to get the final transposition. As suggested by their names, these functions unpack and interleave groups of 64 bits from the low and high halves, respectively, of 128 portions of parameters $a$ and $b$. We note that these functions are defined for floating point parameters but they work well on integers. While casting is needed in C++, no casting instructions are generated in the assembly code. This makes the transpose codes in AVX and SVE similar in the number of machine instructions, as we shall see below.

\texttt{MDSPartial()} firstly computes a dot product between the sponge and a constant vector ($vec[0]$) to determine the new value of the first sponge element ($state[0]$). This is done via the \texttt{spmv\_4x12()} basic function. Next, a vector $W$ is computed as the product between scalar $state[0]$ (note that this is the initial value of $state[0]$) and a constant vector ($vec[1]$). This operation is done with \texttt{mult\_128()} and \texttt{reduce\_128\_64()} basic functions. Vector $W$ is added to the sponge via \texttt{add()}.

\begin{figure}[tp]
\centering
\includegraphics[width=0.65\textwidth]{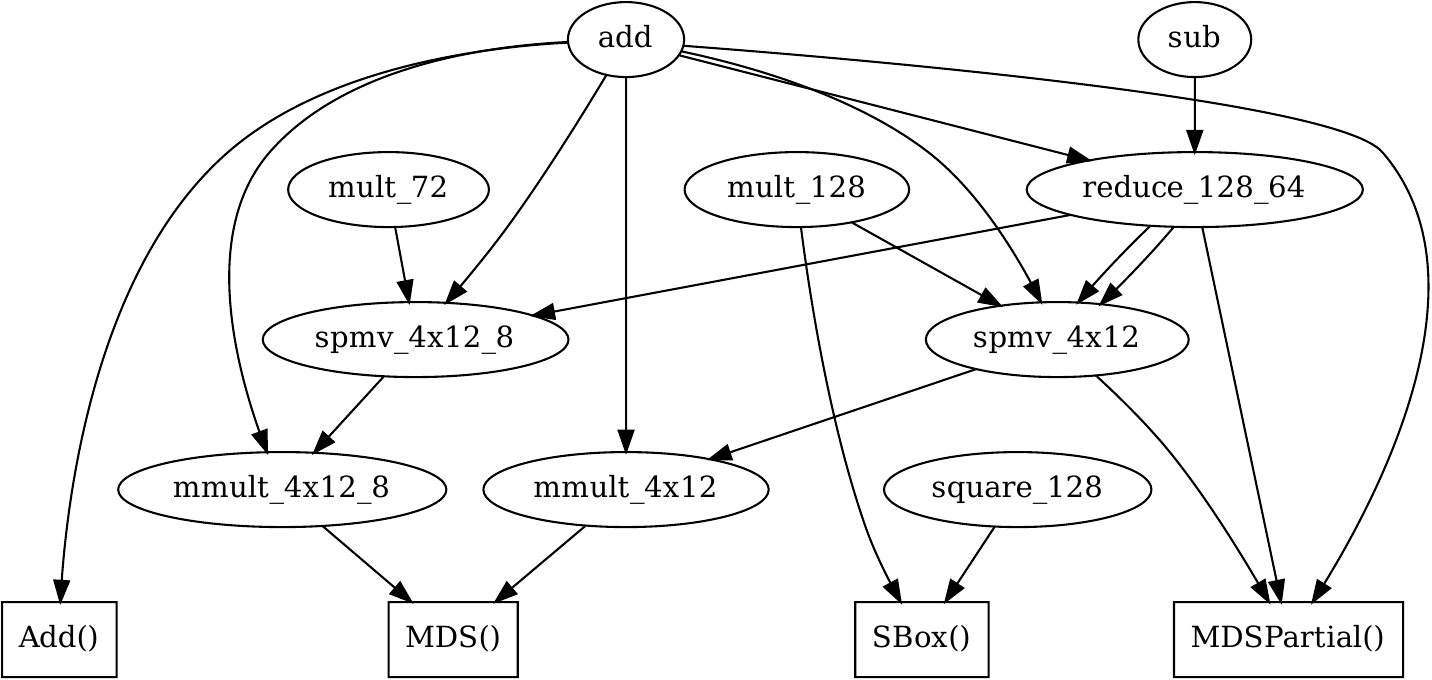}
\caption{Function dependencies in Poseidon implementation.}
\label{fig:graph}
\end{figure}

Finally, we summarize the basic functions called during a Poseidon hash computation in Table~\ref{tbl:funct} and visualize their dependencies in Fig.~\ref{fig:graph}. For example, \texttt{SBox()} uses \texttt{square\_128()} and \texttt{mult\_128()}, while \texttt{MDS()} uses all the basic functions via calls to \texttt{mmult\_4x12()} and \texttt{mmult\_4x12\_8()}.

\subsection{Our SVE Implementation}
\label{sec:sveimpl}

As mentioned above, we implement \texttt{Add()} and \texttt{Sub()} directly using unsigned integers in SVE to avoid the shift operations used by the AVX implementation. Moreover, all our operations produce canonical Goldilocks elements, hence, we do not need to apply \texttt{to\_canonical()} functions. The pseudo-code of \texttt{Add()} and \texttt{Sub()} are shown in Algorithms~\ref{alg:addsve} and~\ref{alg:subsve}, respectively. The SVE assembly for Algorithms~\ref{alg:addsve} is shown in Listing~\ref{lstaddsve}.

\begin{figure}
\centering
\begin{minipage}{0.42\textwidth}
\begin{lstlisting}[caption={Difference between SVE2 and SVE implementations of $mul\_128(a, b)$},label=lstmulsve,basicstyle=\ttfamily\footnotesize,numbers=left]
#ifdef __USE_SVE2__
c_l = svmul_u64_z(svptrue_b64(), a, b);
c_h = svmulh_u64_z(svptrue_b64(), a, b);
#else
// Split into 32 bits
a_h = svlsr_n_u64_z(svptrue_b64(), a, 32);
b_h = svlsr_n_u64_z(svptrue_b64(), b, 32);
a_l = svand_u64_z(svptrue_b64(), a, LMASK);
b_l = svand_u64_z(svptrue_b64(), b, LMASK);
c_hh = svmul_u64_z(svptrue_b64(), a_h, b_h);
c_hl = svmul_u64_z(svptrue_b64(), a_h, b_l);
c_lh = svmul_u64_z(svptrue_b64(), a_l, b_h);
c_ll = svmul_u64_z(svptrue_b64(), a_l, b_l);
c_ll_h = svlsr_n_u64_z(svptrue_b64(), c_ll, 32);
r0 = svadd_u64_z(svptrue_b64(), c_hl, c_ll_h);
r0_l = svand_u64_z(svptrue_b64(), r0, P_n);
r1 = svadd_u64_z(svptrue_b64(), c_lh, r0_l);
r1_l = svlsl_n_u64_z(svptrue_b64(), r1, 32);
sel = svdupq_n_b32(0, 1, 0, 1);
c_l = svreinterpret_u64(svsel_u32(sel, 
    svreinterpret_u32(r1_l), 
    svreinterpret_u32(c_ll)));
r0_h = svlsr_n_u64_z(svptrue_b64(), r0, 32);
r2 = svadd_u64_z(svptrue_b64(), c_hh, r0_h);
r1_h = svlsr_n_u64_z(svptrue_b64(), r1, 32);
c_h = svadd_u64_z(svptrue_b64(), r2, r1_h);
#endif
\end{lstlisting}
\end{minipage}
\end{figure}

\begin{table*}
\centering
\caption{Key function in Poseidon Hash with Goldilocks Field.}
\label{tbl:funct}
\resizebox{0.98\textwidth}{!}{
\begin{tabular}{r|l}
    \textbf{Function Name} & \multicolumn{1}{c}{\textbf{Description}} \\
     \hline
    $add$ & Adds two Goldilocks elements and produces a Goldilocks field element. \\
    $sub$ & Subtracts two Goldilocks elements and produces a Goldilocks field element. \\
    $reduce\_128\_64$ & Reduces a 128-bit number to a (64-bit) Goldilocks field element. \\
    $square\_128$ & Computes the square of a Goldilocks field element. The result is reduced to a Goldilocks field element. \\    
    $mult\_128$ & Computes the product of two Goldilocks field elements. The result is reduced to a Goldilocks field element. \\
    \multirow{2}{*}{$mult\_72$} & Computes the product of two Goldilocks field elements where one of the elements has only 8 bits. \\
    & The result is reduced to a Goldilocks field element. \\
    $spmv\_4x12$ & Computes a sparse matrix-vector product (where the matrix is 4x12). \\
    $spmv\_4x12\_8$ & Computes a sparse matrix-vector product (where the matrix is 4x12 and the vector has 8-bit elements).\\
    $mmult\_4x12$ & Computes a dense matrix-vector product. \\
    $mmult\_4x12\_8$ & Computes a dense matrix-vector product (where the vector has 8-bit elements).\\
    \hline
\end{tabular}
}
\end{table*}

In Algorithm~\ref{alg:addsve}, $c = a + b$ may overflow (both over $P$ or $2^{64} - 1$), while $d = P - a$ will not underflow because $0 \leq a < P$. $c$ overflows only when $b \geq d$; in this case, we need to make it canonical by subtracting $P$ (note that $c$ cannot be greater than $2P$) and we have two possibilities. First, $P \leq c < 2^{64}$, in which case the subtraction does not underflow and the final result is $0 \leq r < 2^{64} - P$. Second, $2^{64} \leq c < 2P$ and $a + b$ overflows. In this case, $c$ is represented on 64 bits as $c^{\prime} = c - 2^{64}$, where $0 \leq c^{\prime} < 2P - 2^{64}$. By subtracting $P$, we have an underflow in the form $-P \leq c^{\prime} - P < P - 2^{64}$. With unsigned integers, a negative value $-P$ is represented as $2^{64} - P$. Hence, $2^{64} - P \leq c^{\prime} - P + 2^{64} < P$ which in the end becomes $2^{64} - P \leq r < P$. This shows that Algorithm~\ref{alg:addsve} produces a canonical Goldilocks element for the addition of two canonical Goldilocks.

In Algorithm~\ref{alg:subsve}, $c = a - b$ will underflow if $a < b$. In this case, $c$ is represented as $c^{\prime} = 2^{64} + a - b$ on a 64-bit unsigned integer, where $2^{64} + 1 - P \leq c^{\prime} < 2^{64}$. Adding $P$ to this produces an overflow but everything above $2^{64}$ is ignored, hence, $1 \leq c^{\prime} + P < P$. In the end, the result is a canonical Goldilocks element.

Next, we discuss the SVE transpose code as shown in Fig.~\ref{fig:transsve}. Similar to AVX, we perform two steps, first to get some intermediary registers and second to get the final transposition. This is done with \texttt{svzip1\_u64(a,b)} and \texttt{svzip2\_u64(a,b)} SVE instructions. \texttt{svzip1\_u64(a,b)} extracts 64 bits from the lowest-indexed halves of the operands and interleaves them (that is, 64 bits from $a$ are followed by 64 bits from $b$). In contrast, \texttt{svzip2\_u64(a,b)} extracts 64 bits from the highest-indexed halves of the operands. In conclusion, the actual operations are different between the AVX and SVE implementations for transpose, but the numbers of machine operations generated by the compiler are the same.

Finally, we show how the newer SVE2 ISA helps to significantly reduce the number of instructions needed to implement the multiplication of two Goldilocks elements. The first step of this multiplication is to obtain the 128-bit product $r = a \cdot b$ of two Goldilocks elements $a$ and $b$. There is no SVE (and for that matter, no AVX/AVX512) instruction to compute this 128-bit product. Instead, there is the \texttt{svmul\_u64\_z()} instruction that returns the low half (64 bits) of the product. The workaround is to split the operands into 32-bit halves, perform four multiplications, and then add the corresponding factors to obtain the 64-bit product halves, as shown under the \textit{else} branch of code Listing~\ref{lstmulsve}. In SVE2, there is the \texttt{svmulh\_u64\_z()} instruction that returns the high half of a 128-bit product. Hence, we only need two multiplications to obtain the product halves, as shown in lines 2-3 of Listing~\ref{lstmulsve}.

\subsection{SVE Code}
The AVX implementation is done in C/C++ using intrinsics (i.e., built-in functions)~\cite{mulavx} and the code is compiled with \texttt{g++} from the GNU C Compiler (GCC). We adopted the same approach for the ARM SVE implementation but, surprisingly, the generated code performs poorly when using GCC-12 or GCC-13. After investigating the assembly code, we discovered that GCC-13 either generates useless instructions (such as \texttt{movprfx}) or has an inefficient register allocation strategy. We show in Listing~\ref{lstaddsvegcc} the assembly generated by GCC-13 for the Goldilocks \texttt{Add()} function presented in Algorithm~\ref{alg:addsve}. In contrast, Listing~\ref{lstaddsve} shows our hand-written SVE assembly. Ignoring the load/store instructions, our hand-written code has seven instructions, while the code generated by GCC has eleven instructions. We observe that a few \texttt{movprfx} instructions are generated to save the content of SVE registers and a \texttt{sel} instruction is generated to get the final result. These are not needed. Firstly, with a proper register allocation, saving SVE registers is not needed. Secondly, a predicated \texttt{add} (as shown on line 9 of Listing~\ref{lstaddsve}) is sufficient to produce the final result (in contrast, GCC generates both an \texttt{add} and a \texttt{sel}). 

On the other hand, GCC-14 generates better machine code. We evaluated other compilers that support ARM SVE, such as \texttt{CLANG} (v18) and ARM Compiler for Linux \texttt{ARMCLANG} (v24.04)~\cite{armcc}. On average, GCC-14 with no hand-written assembly achieves the best performance, while GCC-13 with no hand-written assembly exhibits the worst performance. Hence, in the remainder of this paper, we use GCC-14.

\section{Performance Analysis}
\label{sec:perf}

In this section, we compare the performance of our SVE and SVE2 implementations with the existing AVX and AVX512 implementations on cloud-based instances. We first present the details of these instances, followed by the overall performance when building a Merkle tree. We end the evaluation with a deep dive into the performance metrics of the software under evaluation.

\subsection{Cloud Servers Under Evaluation}
\label{sec:sys}


ARM SVE is represented by four different CPUs deployed in the cloud, namely Graviton3, Graviton4, Yitian 710, and Axion. Graviton3~\cite{graviton3} and Graviton4~\cite{graviton4} CPUs are from \textbf{c7g} and \textbf{c8g} AWS instance families, respectively (in particular, $c7g.metal$ and $c8g.metal-24xl$ instance types). $c7g.metal$ has a Graviton3 CPU with 64 cores clocked at 2.6~GHz and 128~GB of DDR5 RAM. This CPU is based on Neoverse V1 architecture~\cite{v1diag} and it implements the ARMv8.6-a ISA with SVE support where the SVE registers have 256 bits. $c8g.metal-24xl$ has a Graviton4 CPU with 96 cores clocked at 2.8~GHz and 768~GB of DDR5 RAM. This CPU is based on Neoverse V2 architecture~\cite{v2diag} and it implements the ARMv9-a ISA with SVE2 support where the registers have 128 bits. Yitian 710~\cite{yitian1} is a CPU  developed by Alibaba Cloud that powers the \textbf{c8y} VM family. In particular, we use the \textit{ecs.c8y.16xlarge} instance with 64 vCPUs and 128~GB DDR5 from Alibaba Cloud. The Yitian 710~\cite{yitian1} CPU has 64 cores and it is clocked at 3~GHz based on our measurements. This CPU implements ARM-v9 ISA, being based on the Neoverse N2 architecture~\cite{neoversen2}, and supports SVE2 with 128-bit registers. Axion~\cite{introaxion} CPU powers the \textbf{c4a} VM family from GCP. This CPU has a maximum of 72 cores, and based on our estimations\footnote{There is no official statement regarding Axion's clock frequency. Moreover, the Linux \texttt{perf} tool does not work on the c4a instances from GCP (performance monitoring is disabled).}, it is clocked at 3.1~GHz. Similar to Graviton4, this CPU implements the Neoverse V2 architecture with ARM-v9 ISA and supports SVE2 with 128-bit registers. To represent Axion, we select the \textit{c4a-standard-72} VM from GCP which has 72 vCPUs and 288 GB DDR5 memory.

AVX/AVX512 is represented by two CPUs that equip AWS cloud instances, namely \textbf{c7i} and \textbf{c7a} AWS instance families. A \textit{c7i.metal-48xl} is equipped with 192 cores of Intel Xeon Scalable 8488C type clocked at a base frequency of 3.2~GHz and 384~GB DDR5 RAM. We note that the frequency is increased to 3.7~GHz by the Intel Turbo Boost technology based on our measurements with the Linux \texttt{perf} tool. A \textit{c7a.metal-48xl} has 192 cores of AMD EPYC 9R14 type clocked at 3.7~GHz and 384~GB DDR5 RAM. 

We summarize the cloud instances under test in Table~\ref{table:sys_char}. All these instances run the Ubuntu 24.04 LTS operating system. We use GCC 14.2 (\texttt{g++}) to compile the code and \texttt{perf} to get performance metrics. All the experiments are run at least five times and we present the average in this paper. The standard deviation is less than $4\%$ of the average for all the results reported in this paper. 

\subsection{Performance of Merkle Tree Building}
\label{sec:mtperf}

We first analyze the performance of building a Merkle tree with $2^{22}$ leaves where each leaf consists of eight Goldilocks elements (that is, each leaf has 64 bytes). We choose this number because a Poseidon permutation is done on eight input elements. With a leaf of eight Goldilocks, we have one Poseidon permutation per hash. Similarly, the hash of a concatenation of two other hashes operates on eight Goldilocks. 
 
Merkle tree building accounts for 30\% of \prover execution time. When building a Merkle tree, we first compute the Poseidon hashes of the leaves in parallel, since there is no data dependency among these hashes. Then, we proceed level by level by computing hashes of the concatenation of two hashes on the previous level (children hashes). Such concatenation has 64 bytes (eight Goldilocks elements). We end the process once we compute the hash representing the root of the tree.

\begin{figure}[tp]
\centering
\includegraphics[width=0.65\textwidth]{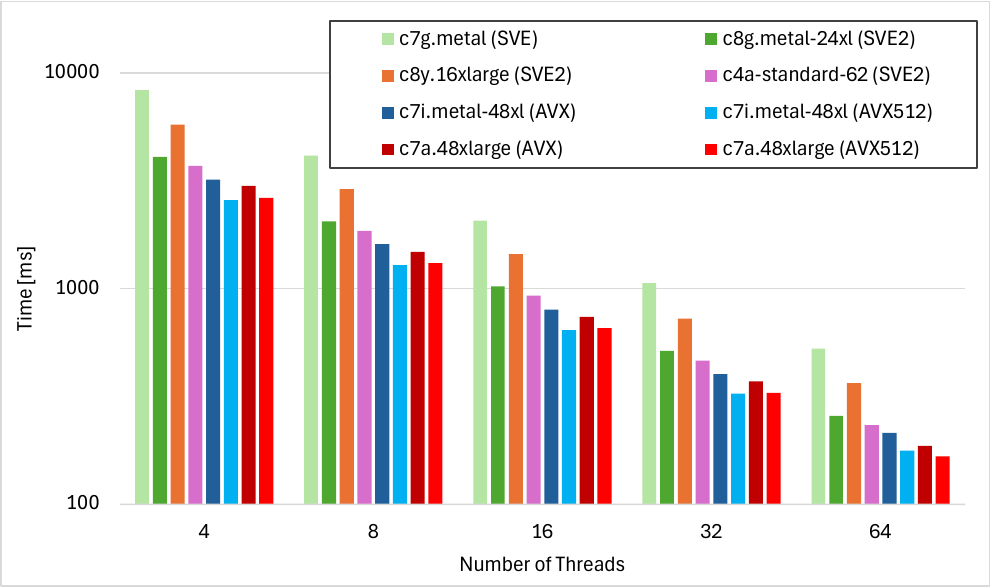}
\caption{Comparison of the execution times for Merkle tree building on different CPUs (logarithmic scale).}
\label{fig:mt}
\end{figure}

We compare our SVE and SVE2 implementations running on \gthree, \gfour, \yitian, and \axion with the AVX and AVX512 implementations running on \amd and \xeon. As shown in Fig.~\ref{fig:mt}, Xeon and EPYC CPUs with AVX512 exhibit the best (lowest) execution time, followed by AVX. SVE, represented by \gthree, is far behind, being $2.6\times$ and $2.8\times$ slower than the AVX implementation running on c7i and c7a, respectively. Furthermore, the SVE implementation is $3.2\times$ slower compared to the AVX512 implementation on both c7i and c7a. Another limitation of c7g is its low number of cores (only 64 cores). In contrast, c7i and c7a each have 192 cores, being able to scale well on large Merkle trees.

\begin{table*}[t]
\centering
\caption{Specifications of the servers under test}
\label{table:sys_char}
\resizebox{\textwidth}{!} {
\begin{tabular}{|l|r|r|r|r|r|r|}
\hline
& \textbf{\gthree} & \textbf{\gfour} & \textbf{\yitian} & \textbf{\axion} & \textbf{\amd} & \textbf{\xeon} \\
\hline
ISA                  & ARMv8.4 & ARMv9 & ARMv9 & ARM v9 & x86-64   & x86-64 \\
Micro-architecture         & Neoverse V1 & Neoverse V2 & Neoverse N2 & Neoverse V2 & Genoa (Zen 4) & Golden Cove \\
Maximum Cores (per CPU) & 64 & 96 & 128 & 72 & 96 (192) & 48 (96) \\
Maximum (v)Cores (per VM)        & 64      & 192      & 128   & 72   &  192    & 192 \\
Clock Frequency      & 2.6~GHz & 2.8~GHz & 3~GHz & 3.1~GHz & 3.7~GHz & 3.2/3.7~GHz \\
\hline
\end{tabular}
}
\end{table*}

SVE2, represented by \gfour, \yitian, and \axion, exhibits better performance compared to SVE represented by \gthree. Among the three CPUs that support SVE2, \axion exhibits the fastest execution time, being $9.4\%$ and $36\%$ faster than \gfour and \yitian, respectively. However, \axion is still $1.4\times$ slower compared to \xeon and \amd using AVX512. We note that \gfour and \axion both implement the Neoverse V2 microarchitecture, but Axion archives higher performance. We attribute this to a higher clock frequency. We estimate that the CPU clock frequency in a \textit{c4a} VM is 3.1~GHz, which is $9.4\%$ higher than the 2.8~GHz clock frequency of \gfour.

As a side note, we observe that \amd is faster than \xeon when running the AVX implementation but both CPUs achieve the same performance when running the AVX512 implementation. We attribute this to the Intel Turbo Boost technology, which does not increase the clock frequency to a maximum of 3.7~GHz when AVX functions are executed. For example, our profiling with the Linux tool \texttt{perf} shows that the CPU clock frequency of the \xeon instance is 3.4 - 3.6~GHz during the execution of \texttt{add()} and \texttt{sub()} functions with AVX.

\subsection{Performance of Basic Functions}

\begin{figure}[tbp]
\centering
\includegraphics[width=0.65\textwidth]{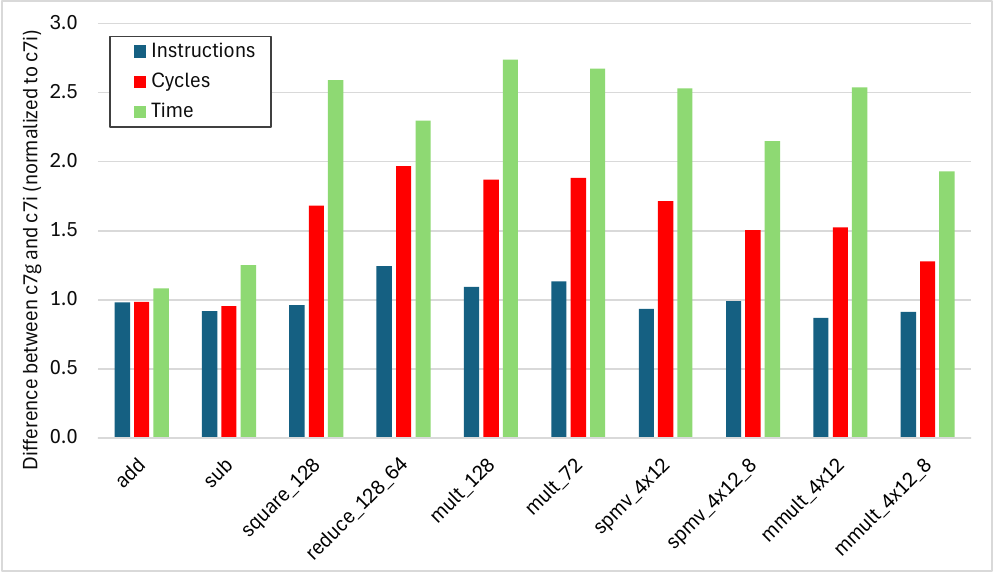}
\caption{Comparison of performance metrics (instructions, cycles, time) between \gthree and \xeon.}
\label{fig:ict}
\end{figure}

\textbf{SVE vs. AVX.} We analyze the performance of the basic functions as listed in Table~\ref{tbl:funct}. We present the metrics of \gthree normalized to those of \xeon in Fig.~\ref{fig:ict}. These metrics are the number of instructions, the number of cycles, and execution time, and they are obtained with the Linux \texttt{perf} tool which monitors hardware-based counters.

We first observe that our SVE implementation renders similar or even lower (e.g., for \texttt{add()} and \texttt{sub()}) numbers of instructions compared to AVX. In the case of \texttt{add()} and \texttt{sub()}, our SVE implementation uses fewer instructions compared to AVX/AVX512, as discussed in Section~\ref{sec:impl}. For the other functions, we use instructions similar to those available in AVX/AVX512.
However, the number of cycles is more than $1.5\times$ higher on \gthree, except for \texttt{add()}, \texttt{sub()}, and \texttt{mmult\_4x12\_8()}. For the first two functions, the number of cycles needed by \gthree is slightly lower compared to \amd and \xeon. These two functions mainly rely on vectorized addition and subtraction instructions, while all other basic functions include multiplication instructions. To get a better picture, we benchmark the performance of vectorized addition, subtraction, and multiplication instructions in isolation for both ARM SVE and x86-64 AVX. Addition / subtraction exhibits 2.4 and 4.5 instructions-per-cycle (IPC) on \gthree (SVE) and \xeon (AVX), respectively. These numbers are in correlation with the number of vector execution units in \gthree and \xeon, which are two and four, respectively, and the fact that addition and subtraction require one cycle to complete. On the other hand, multiplication shows an IPC of 0.6 and 1.75 in \gthree (SVE) and \xeon (AVX), respectively. For \xeon, this number is similar to the IPC reported by Intel on their website~\cite{mulavx}. But for \gthree, the IPC is very low, suggesting that multiplication needs four cycles to complete (that is, even if there are two vector execution units, these units can complete two multiplications in four cycles, leading to an IPC of 0.5). This low multiplication performance and the low number of vector execution units in \gthree lead to worse performance compared to \amd and \xeon.

Finally, the trend of the execution time gap between \gthree and \xeon follows that of the cycles. This is as expected since \gthree has a clock frequency that is $1.4\times$ lower than \xeon and \amd (that is, 2.6~GHz vs. 3.7~GHz). In summary, at yet another level, the current state-of-the-art ARM SVE hardware is far behind its AVX counterparts. 

\begin{figure}[tbp]
\centering
\includegraphics[width=0.65\textwidth]{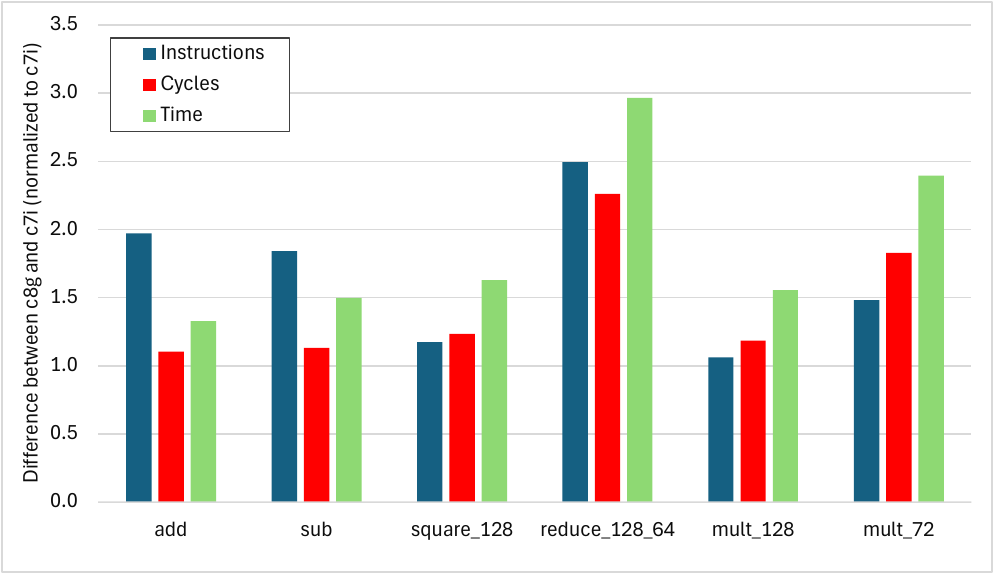}
\caption{Comparison of performance metrics (instructions, cycles, time) between \gfour and \xeon.}
\label{fig:ict-sve2}
\end{figure}

\textbf{SVE2 vs. AVX.} Next, we analyze the performance of some basic functions implemented with SVE2 and 128-bit registers compared to their AVX 256-bit counterparts. For this analysis, we compare the performance metrics collected on \gfour to those of \xeon. As shown in Fig.~\ref{fig:ict-sve2}, we analyze the functions that can be directly compared, such as \texttt{add()}, \texttt{sub()}, \texttt{square\_128()}, \texttt{reduce\_128()}, \texttt{mult\_128()}, and \texttt{mult\_72()}. The other four functions, namely \texttt{spmv\_4x12()}, \texttt{spmv\_4x12\_8()}, \texttt{mmult\_4x12()}, and \texttt{mmult\_4x12\_8()} have different implementations when the target registers are 128-bit (such as our SVE2 target CPUs, namely \gfour, \yitian, and \axion) compared to 256-bit (such as our SVE target CPU, \gthree). Hence, it is not fair to compare these four functions to their 256-bit AVX counterparts. 
 
For the six selected functions, to ensure a fair comparison, we run the benchmarks on inputs of the same length as for AVX. For example, if the input has 64 bytes (256 bits), we run the AVX \texttt{add()} once, while the SVE2 \texttt{add()} is run twice, once for each half of the 256-bit value. Thus, the number of instructions needed by the SVE2 implementations is almost double for \texttt{add()} and \texttt{sub()}, as shown in Fig.~\ref{fig:ict-sve2}. It is not exactly double because our SVE2 implementations use fewer instructions compared to AVX, as explained in Section~\ref{sec:impl}. In \texttt{square\_128()}, \texttt{mult\_128()}, and \texttt{mult\_72()} we make use of SVE2's \texttt{svmulh\_u64\_z()} instruction which returns the high half of a 128-bit result of two 64-bit operands multiplication, as detailed in Section~\ref{sec:sveimpl}. This significantly reduces the number of instructions needed, so that even if we execute them twice (once for each 128-bit part of a 256-bit target), the total number of instructions needed by the SVE2 implementations is lower compared to AVX. On the other hand, we observe that these functions exhibit lower IPC compared to \texttt{add()} and \texttt{sub()}, because they utilize multiplication instructions that have higher latency compared to addition. The \texttt{reduce\_128\_64()} function exhibits significantly worse performance compared to the other functions: it needs $2.5\times$ more instructions compared to its AVX counterpart. This is because a single instance of this function (applied on 128 bits) consists of 17 instructions (hence, 34 instructions when applied on 256 bits), while its AVX counterpart (applied on 256 bits) consists of 14 instructions. 

In summary, we find that SVE2 ISA is more powerful compared to AVX and AVX512 (e.g., it supports operations with unsigned integers and has instructions that allow us to get both the high and low parts of 128-bit product results of 64-bit operand multiplication). On the other hand, the current hardware implementations have only 128-bit registers and the latency of multiplications is higher in the evaluated ARM CPUs compared to their x86-64 counterparts.

\section{Discussion}
\label{sec:dis}

In this section, we perform a ``what-if'' analysis to gauge the impact on execution time, power, and price of improving the performance of ARM CPUs with SVE2 support. In particular, we are looking at the effect of (1) increasing the clock frequency of ARM CPUs to match their x86-64 counterparts and (2) increasing the size of SVE2 registers and execution units from 128 bits to 256 and 512 bits. In this analysis, we use \gfour and \axion as reference ARM CPUs. Moreover, we use the following execution time model: given a bunch of instructions $I$, the corresponding number of cycles $C$ to run these instructions, and the clock frequency $f$ of the CPU, the time $T$ to run these instructions is 
\begin{equation}
\label{eq:ct}
T = \frac{C}{f} = \frac{I}{IPC \cdot f}
\end{equation}
When the frequency is increased, the number of instructions and the number of cycles remain unchanged. Hence, a multiplication factor $m$ for the frequency ($f' = m \cdot f$) results in an execution time reduced by $m$ times ($T' = \frac{T}{m}$).

When increasing the size of the SVE2 registers and execution units, while keeping the input data size constant, the number of instructions needed to compute a Poseidon hash decreases because we operate on larger data chunks with each SVE2 instruction. Since there is no available ARM CPU with 256- or 512-bit SVE2 registers, we use the QEMU~\cite{qemu} emulator to count the number of 64-bit ARM instructions emulated on an x86-64 host when building Merkle trees. The results are consistent across different Merkle tree sizes: 256- and 512-bit SVE2 need $45.7\%$ and $54.7\%$, respectively, fewer instructions compared to 128-bit SVE2. 

\begin{figure}[tbp]
\centering
\includegraphics[width=0.65\textwidth]{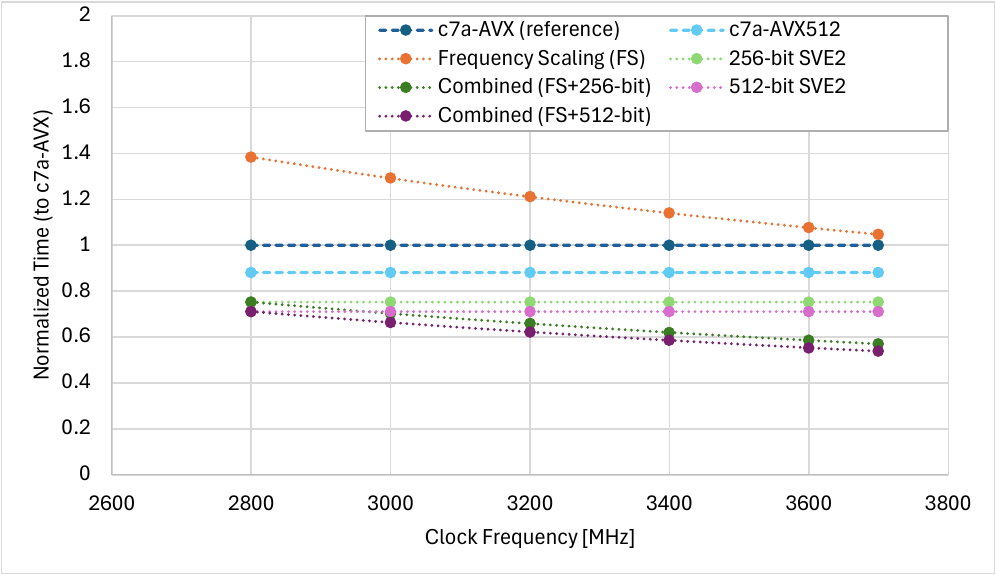}
\caption{Normalized execution times for Merkle tree building with 64 threads on \gfour when the clock frequency is scaled up to 3.7 GHz and the SVE2 register size is increased to 256 bits (normalized to the execution time on \amd).}
\label{fig:mt-freq}
\end{figure}

As shown in Fig.~\ref{fig:mt-freq}, only scaling the frequency does not make the performance of an ARM CPU with SVE2 better than that of Xeon and EPYC CPUs with AVX. On the other hand, if we increase the register size of the SVE2 registers and execution units to 256 bits, even if we run at the original frequency of 2.8~GHz, the ARM CPU is around $25\%$ faster than \amd with AVX. If we combine this increase in register size with frequency scaling, we achieve a total improvement of approximately $43\%$ compared to \amd when the frequency is 3.7~GHz. Increasing the register size to 512 bits results in $29\%$ faster execution time on ARM compared to \amd with AVX512. When combining this with frequency scaling, the maximum performance gain on ARM compared to \amd is $46\%$.

Although these results are encouraging, we also need to consider the impact on power and cost when increasing hardware performance. For example, we can estimate a linear increase in power with frequency based on the dynamic switching power of a digital circuit~\cite{edcirc}: $P_{d} = C \cdot V^2 \cdot f$, where $C$ is the capacitance of the circuit, $V$ is the operating voltage, and $f$ is the clock frequency. In reality, frequency and voltage are interrelated, but we leave the design of a more accurate power model for future research. Estimating the impact on power of increasing the SVE register size is even more complex and becomes impossible without access to the hardware design. Hence, we make the pessimistic assumption that increasing both the frequency and the size of the SVE2 register leads to an $2\times$ increase in dynamic power. Next, based on our vast experience in measuring the performance and power of modern CPUs~\cite{LOGHIN20191,dumi_edge2017}, we assume that dynamic power accounts for $80\%$ of the total power used by a CPU. Furthermore, we make the following assumptions which are backed by data shared by Google in the third edition of ``The Datacenter as a Computer'' book~\cite{GoogleDatacenter_19}: the CPU accounts for $61\%$ of the power usage of a server and the servers' power accounts for $5-12\%$ of the total cost of ownership of a datacenter~\cite{GoogleDatacenter_19}. Putting everything together, we have the initial power usage of a CPU:
\begin{equation}
    P_{CPU} = P_s + P_d, P_s = 0.2\cdot P_{CPU}, P_d = 0.8\cdot P_{CPU}
\end{equation}
Increasing the performance results in $2\times$ dynamic power, 
\begin{equation}
    P^{'}_{CPU} = 0.2\cdot P_{CPU} + 2\cdot 0.8\cdot P_{CPU} = 1.8 \cdot P_{CPU}
\end{equation}
At the datacenter level, the new power usage is:
\begin{equation}
    P^{'} = 0.39\cdot P + 0.61\cdot 1.8\cdot P = 1.488 \cdot P
\end{equation}
Using $\gamma$ to denote the proportion of energy costs in the total cost of ownership of a datacenter, and the fact that energy and power are directly proportional, we obtain the following new total cost of ownership,
\begin{equation}
    C^{'} = (1-\gamma)\cdot C + \gamma \cdot 1.488 \cdot C
\end{equation}
When $\gamma$ is in the range $5-20\%$, the total cost increment is in the range $2.44-9.76\%$. This is an acceptable increase in price. For example, the \textit{c7i} and \textit{c7a} instances used in our experimental evaluation cost $11.9\%$ and $28.7\%$, respectively, more than the \textit{c8g} instance. Considering the performance gains and the maximum increase in cost of $10\%$, ARM instances will conserve their competitive price advantage.


\section{Related Work}

To our knowledge, this is the first study to combine ARM SVE, ZKP, and cloud computing. Other papers that look at SVE do it in the context of HPC on supercomputers~\cite{fugaku, mpisve, hpcsve, svecluster} and do not analyze the performance of ZKP primitives. We note that it is not feasible to run the \prover on a supercomputer since this is part of a highly available online service. Our previous study of cloud-based ARM servers~\cite{dumi_tcc2024} does not focus on SVE since none of the database workloads analyzed in that study are implemented with SVE support. In contrast, we focus exclusively on ARM SVE in this paper by implementing the SVE and SVE2 versions of Goldilocks field operations used by \prover. In fact, this is the first use of ARM SVE/SVE2 in ZKP so far. Moreover, this is the first performance study including Graviton4 and Axion CPUs.

\section{Conclusions}

In this paper, we presented and evaluated our ARM SVE/SVE2 implementations of Goldilocks field operations in the context of Merkle tree building with Poseidon hashing which is a key component of ZK proving systems, such as \prover~\cite{zkprover_doc}. We compared our implementation with existing AVX and AVX512 implementations that run on Intel and AMD processors. Our performance evaluation is conducted on cloud-based instances with ARM, Intel, and AMD CPUs.
From the point of view of ARM CPUs, our study brings both good and bad news. On the positive side, we observe that the SVE/SVE2 ISA is powerful enough to efficiently implement all the Goldilocks field operations. In fact, by supporting unsigned integers, it is easier to implement Goldilocks addition and subtraction in SVE compared to AVX, resulting in fewer machine instructions per operation. On the negative side, the current hardware implementations of ARM SVE/SVE2 are far behind their x86-64 (Intel and AMD) counterparts. Current x86-64 servers have hundreds of cores (e.g., 192 in our evaluation), while only the \gfour cloud instances offer a maximum of 192 ARM cores in one VM. SVE2 ARM CPUs only support 128-bit vectors, compared to 512-bit vectors in AVX512 CPUs. In addition, the clock frequency of x86-64 CPUs is higher (e.g., 3.7~GHz for the Intel and AMD CPUs in our study compared to 2.8, 3.1, and 3~GHz for \gfour, \axion, and \yitian, respectively). All these limitations lead to $1.4-3\times$ lower performance of current ARM CPUs compared to their x86-64 counterparts. However, when ARM CPUs will implement 512-bit SVE2 execution units, we expect their performance to surpass that of current x86-64 CPUs with AVX512.



\begin{thebibliography}{10}

\bibitem{aws_arm}
AWS, ``{AWS Graviton Processor}.'' https://archive.ph/VFtpU, 2023.

\bibitem{dumi_tcc2024}
D.~Loghin, ``{Are ARM Cloud Servers Ready for Database Workloads? An
  Experimental Study},'' {\em IEEE Transactions on Cloud Computing}, pp.~1--12,
  2024.

\bibitem{svepaper}
N.~Stephens, S.~Biles, M.~Boettcher, J.~Eapen, M.~Eyole, G.~Gabrielli,
  M.~Horsnell, G.~Magklis, A.~Martinez, N.~Premillieu, A.~Reid, A.~Rico, and
  P.~Walker, ``{The ARM Scalable Vector Extension},'' {\em IEEE Micro},
  vol.~37, no.~2, pp.~26--39, 2017.

\bibitem{poseidon_hash}
L.~Grassi, D.~Khovratovich, C.~Rechberger, A.~Roy, and M.~Schofnegger,
  ``{Poseidon: A New Hash Function for Zero-Knowledge Proof Systems}.''
  Cryptology ePrint Archive, Paper 2019/458, 2019.
\newblock \url{https://eprint.iacr.org/2019/458}.

\bibitem{goldilocks}
R.~Bloemen, ``{The Goldilocks Prime}.'' https://archive.ph/EMqmM, 2024.

\bibitem{zkprover_doc}
Polygon, ``{zkProver}.'' https://archive.ph/K7QMD, 2024.

\bibitem{zk_okx}
OKX, ``What is zk technology? how zero-knowledge boosts blockchain security and
  scalability.'' https://archive.ph/N0dY4, 2024.

\bibitem{zkevm_doc}
Polygon, ``{Polygon zkEVM}.'' https://archive.ph/iripJ, 2024.

\bibitem{xlayer}
OKX, ``{Introduction to X Layer}.'' https://archive.ph/2aEyt, 2024.

\bibitem{flynn}
M.~Flynn, ``{Very High-speed Computing Systems},'' {\em Proc. of IEEE},
  vol.~54, no.~12, pp.~1901--1909, 1966.

\bibitem{avx_book}
D.~Kusswurm, {\em {Modern Parallel Programming with C++ and Assembly Language:
  X86 SIMD Development Using AVX, AVX2, and AVX-512}}.
\newblock Apress, 2022.

\bibitem{introsve}
ARM, ``{Introducing SVE}.'' https://archive.ph/T2Us8, 2024.

\bibitem{introsve2}
ARM, ``{Introducing SVE2}.'' https://archive.ph/muCM8, 2024.

\bibitem{Ethereum_2013}
V.~Buterin, ``{A Next-Generation Smart Contract and Decentralized Application
  Platform}.'' http://archive.fo/Sb4qa, 2013.

\bibitem{rpc_book}
P.~Ruan, T.~T.~A. Dinh, D.~Loghin, M.~Zhang, and G.~Chen, {\em {Blockchains -
  Decentralized and Verifiable Data Systems}}.
\newblock Springer Cham, 2022.

\bibitem{Bitcoin_2008}
S.~Nakamoto, ``{Bitcoin: A Peer-to-peer Electronic Cash System}.''
  http://archive.fo/CIl1Y, 2008.

\bibitem{cdk_doc}
Polygon, ``{Polygon CDK}.'' https://archive.ph/K2Jb8, 2024.

\bibitem{zkprover_paper}
H.~Masip-Ardevol, M.~Guzmán-Albiol, J.~Baylina-Melé, and J.~L. Muñoz-Tapia,
  ``{eSTARK: Extending STARKs with Arguments}.'' Cryptology ePrint Archive,
  Paper 2023/474, 2023.
\newblock \url{https://eprint.iacr.org/2023/474}.

\bibitem{gp_blog}
A.~P. Goucher, ``{An Efficient Prime for Number-theoretic Transforms}.''
  https://archive.ph/7NOeG, 2021.

\bibitem{mulavx}
Intel, ``{Intel Intrinsics Guide}.'' https://archive.ph/ZMwY4, 2024.

\bibitem{armcc}
ARM, ``{Arm Compiler for Linux}.'' https://archive.ph/49H1A, 2024.

\bibitem{graviton3}
J.~Barr, ``{New Graviton3-Based General Purpose (m7g) and Memory-Optimized
  (r7g) Amazon EC2 Instances}.'' https://archive.ph/q8vrz, 2023.

\bibitem{graviton4}
TheNextPlatform, ``{AWS Adopts ARM V2 Cores For Expansive Graviton4 Server
  CPU}.'' https://archive.fo/Scz7X, 2023.

\bibitem{v1diag}
ARM, ``{Neoverse V1}.'' https://archive.fo/uzc5l, 2023.

\bibitem{v2diag}
ARM, ``{Neoverse V2}.'' https://archive.fo/pQDxV, 2023.

\bibitem{yitian1}
AlibabaCloud, ``{ECS Yitian Instance Deep Learning Reasoning Performance
  Measurement}.'' https://archive.fo/jrQqp, 2023.

\bibitem{neoversen2}
ARM, ``{AWS Drives Cloud Price/Performance and CPU Silicon Innovation with Arm
  Neoverse}.'' https://archive.ph/i62Fd, 2021.

\bibitem{introaxion}
Google, ``{Introducing Google Axion Processors, our new Arm-based CPUs}.''
  https://archive.ph/jAWpn, 2024.

\bibitem{qemu}
F.~Bellard, ``{QEMU, a Fast and Portable Dynamic Translator},'' in {\em Proc.
  of USENIX Annual Technical Conference}, p.~41, 2005.

\bibitem{edcirc}
J.~M. Rabaey, {\em {Digital Integrated Circuits: a Design Perspective}}.
\newblock USA: Prentice-Hall, Inc., 1996.

\bibitem{LOGHIN20191}
D.~Loghin and Y.~M. Teo, ``{The Time and Energy Efficiency of Modern Multicore
  Systems},'' {\em Parallel Computing}, vol.~86, pp.~1--13, 2019.

\bibitem{dumi_edge2017}
D.~Loghin, L.~Ramapantulu, and Y.~M. Teo, ``{On Understanding Time, Energy and
  Cost Performance of Wimpy Heterogeneous Systems for Edge Computing},'' in
  {\em Proc. of IEEE International Conference on Edge Computing (EDGE)},
  pp.~1--8, 2017.

\bibitem{GoogleDatacenter_19}
L.~A. Barroso, U.~Hoelzle, and P.~Ranganathan, {\em {The Datacenter as a
  Computer: Designing Warehouse-Scale Machines, Third Edition}}.
\newblock Springer Cham, 3rd~ed., 2019.

\bibitem{fugaku}
M.~Sato, ``{The Supercomputer Fugaku and ARM-SVE Enabled A64FX Processor for
  Energy-efficiency and Sustained Application Performance},'' in {\em Proc. of
  19th International Symposium on Parallel and Distributed Computing (ISPDC)},
  pp.~1--5, 2020.

\bibitem{mpisve}
D.~Zhong, P.~Shamis, Q.~Cao, G.~Bosilca, S.~Sumimoto, K.~Miura, and
  J.~Dongarra, ``{Using Arm Scalable Vector Extension to Optimize OPEN MPI},''
  in {\em Proc. of 20th IEEE/ACM International Symposium on Cluster, Cloud and
  Internet Computing (CCGRID)}, pp.~222--231, 2020.

\bibitem{hpcsve}
D.~Yokoyama, B.~Schulze, and F.~Borges, ``{The Survey on ARM Processors for
  HPC},'' in {\em Journal of Supercomputing}, vol.~75, p.~7003–7036, 2019.

\bibitem{svecluster}
Y.~Kodama, T.~Odajima, M.~Matsuda, M.~Tsuji, J.~Lee, and M.~Sato,
  ``{Preliminary Performance Evaluation of Application Kernels Using ARM SVE
  with Multiple Vector Lengths},'' in {\em Proc. of IEEE International
  Conference on Cluster Computing}, pp.~677--684, 2017.

\end{thebibliography}
\end{document}